\documentstyle[sprocl,epsf]{article}
\input{psfig}

\def\etal{{\it et al.}}

\def\sigdbcd{\relax\ifmmode{ \sigma_{_{\rm DBCD}} }\else{ $\sigma_{_{\rm DBCD}}$\ }\fi}

\def\mGeV{\relax\ifmmode{{\rm GeV/c^2}}\else{${\rm GeV/c^2}$\ }\fi}
\def\mMeV{\relax\ifmmode{{\rm MeV/c^2}}\else{${\rm MeV/c^2}$\ }\fi}
\def\pGeV{\relax\ifmmode{{\rm GeV/c}}\else{${\rm GeV/c}$\ }\fi}
\def\pMeV{\relax\ifmmode{{\rm MeV/c}}\else{${\rm MeV/c}$\ }\fi}

\def\thrupork#1{\mathrel{\mathop{#1\!\!\!/}}}

\def\thru#1{\mathrel{\mathop{#1\!\!\!\!/}}}
\def\thrud#1{\mathrel{\mathop{#1\!\!\!\!/}}}

\def\gammu{\relax\ifmmode{\gamma^\mu}\else{$\gamma^\mu$\ }\fi}
\def\Amu{\relax\ifmmode{A_\mu}\else{$A_\mu$\ }\fi}
\def\Anu{\relax\ifmmode{A_\nu}\else{$A_\nu$\ }\fi}
\def\Dnu{\relax\ifmmode{D_\nu}\else{$D_\nu$\ }\fi}
\def\Dmu{\relax\ifmmode{D_\mu}\else{$D_\mu$\ }\fi}
\def\Fsq{\relax\ifmmode{F_{\mu\nu}F^{\mu\nu}}\else{$F_{\mu\nu}F^{\mu\nu}$\ }\fi}
\def\del{\relax\ifmmode{\partial}\else{$\partial$\ }\fi}

\def\delsl{\relax\ifmmode{\thrupork{\partial}}\else{$\thrupork{\partial}$\ }\fi}
\def\Asl{\relax\ifmmode{\thrupork{A}}\else{$\thrupork{A}$\ }\fi}
\def\Bsl{\relax\ifmmode{\thrud{B}}\else{$\thrud{B}$\ }\fi}
\def\Dsl{\relax\ifmmode{\thrud{D}}\else{$\thrud{D}$\ }\fi}
\def\Bsl{\relax\ifmmode{\thrud{B}}\else{$\thrud{B}$\ }\fi}

\def\delmu{\relax\ifmmode{\partial_\mu}\else{$\partial_\mu$\ }\fi}
\def\delnu{\relax\ifmmode{\partial_\nu}\else{$\partial_\nu$\ }\fi}
\def\psib{\relax\ifmmode{{\bar \psi}}\else{${\bar \psi}$\ }\fi}

\def\Bmu{\relax\ifmmode{B_\mu}\else{$B_\mu$\ }\fi}
\def\Bnu{\relax\ifmmode{B_\nu}\else{$B_\nu$\ }\fi}

\def\Wslmu{\relax\ifmmode{\thru{W}_\mu}\else{$\thru{W}_\mu$\ }\fi}
\def\Wmu{\relax\ifmmode{W_\mu}\else{$W_\mu$\ }\fi}
\def\Wslnu{\relax\ifmmode{\thrud{W}_\nu}\else{$\thrud{W}_\nu$\ }\fi}
\def\Wnu{\relax\ifmmode{W_\nu}\else{$W_\nu$\ }\fi}
\def\exp{\relax\ifmmode{{\rm e}}\else{${\rm e}$\ }\fi}

\def\SUiiL{\relax\ifmmode{SU(2)_L}\else{$SU(2)_L$\ }\fi}
\def\SUiiiC{\relax\ifmmode{SU(3)_C}\else{$SU(3)_C$\ }\fi}
\def\SUiif{\relax\ifmmode{SU(2)_f}\else{$SU(2)_f$\ }\fi}
\def\SUiiif{\relax\ifmmode{SU(3)_f}\else{$SU(3)_f$\ }\fi}
\def\SUiis{\relax\ifmmode{SU(2)_s}\else{$SU(2)_s$\ }\fi}
\def\SUivf{\relax\ifmmode{SU(4)_f}\else{$SU(4)_f$\ }\fi}
\def\SUiv{\relax\ifmmode{SU(4)}\else{$SU(4)$\ }\fi}
\def\UiY{\relax\ifmmode{U(1)_Y}\else{$U(1)_Y$\ }\fi}
\def\SUii{\relax\ifmmode{SU(2)}\else{$SU(2)$\ }\fi}
\def\SUiii{\relax\ifmmode{SU(3)}\else{$SU(3)$\ }\fi}
\def\Ui{\relax\ifmmode{U(1)}\else{$U(1)$\ }\fi}

\def\codefont{\tt}
\def\startline{\par\noindent}
{\obeylines\obeyspaces%
\global\def\beginCode{\displaybreak%
\begingroup%
\singlespace%
\parskip=0pt%
\obeylines\obeyspaces%
\let^^M=\startline%
\codefont}%
}

\def\displaybreak{\bigbreak}

\def\jhead#1#2{			
  \global\advance\headnumber by 1
  \updownpages\tocitem{#2}
  {\immediate\write16{#2} 
   \raggedcenter {\bf#1.\the\headnumber \hskip .5cm }{\bf {\sc #2}} \par }
   \nobreak\vskip 0.1truein\nobreak\global\headpage=\pageno\beginparmode}

\def\beginMidItems{\begingroup 
     \vskip 20pt
     \advance\parindent by 8em}

\def\beginLeftItems{\begingroup
     \vskip 20pt
     \advance\parindent by 2em}

\def\sigediff{\relax\ifmmode{\sigma_{\ediff}}\else{$\sigma_{\ediff}$\ }\fi}
\def\Ecm{\relax\ifmmode{E_{cm}}\else{$E_{cm}$\ }\fi}

\def\cleo{CLEO II }

\def\sig{\relax\ifmmode{\sigma}\else{$\sigma$\ }\fi}
\def\degr{\relax\ifmmode{^\circ}\else{$^\circ$}\fi}
\def\ediff{\relax\ifmmode{\Delta{\rm E}}\else{$\Delta{\rm E}$\ }\fi}
\def\bmass{\relax\ifmmode{{\rm M_{BC}}}\else{${\rm M_{BC}}$\ }\fi}
\def\spheri{\relax\ifmmode{\Theta_f}\else
	{$\Theta_f$\ }\fi}
\def\heli{\relax\ifmmode{\Theta_h}\else
	{$\Theta_h$\ }\fi}
\def\dmdiff{\relax\ifmmode{m_{D^{*}}-m_{D}}\else
	{${m_{D^{*}}-m_{D}}$\ }\fi}

\def\parone{\sl}

\def\tiny{\vrule width 0pt}
\def\star{{\bf *}}
\def\epem{\relax\ifmmode{e^+e^-}\else{$e^+e^-$\ }\fi}
\def\decays{\relax\ifmmode{\rightarrow}\else{$\rightarrow$\ }\fi\tiny}

\def\Vud{\relax\ifmmode{{\rm V}_{ud}}\else{{\rm V}$_{ud}$\ }\fi}
\def\Vcd{\relax\ifmmode{{\rm V}_{cd}}\else{{\rm V}$_{cd}$\ }\fi}
\def\Vtd{\relax\ifmmode{{\rm V}_{td}}\else{{\rm V}$_{td}$\ }\fi}
\def\Vus{\relax\ifmmode{{\rm V}_{us}}\else{{\rm V}$_{us}$\ }\fi}
\def\Vcs{\relax\ifmmode{{\rm V}_{cs}}\else{{\rm V}$_{cs}$\ }\fi}
\def\Vts{\relax\ifmmode{{\rm V}_{ts}}\else{{\rm V}$_{ts}$\ }\fi}
\def\Vub{\relax\ifmmode{{\rm V}_{ub}}\else{{\rm V}$_{ub}$\ }\fi}
\def\Vcb{\relax\ifmmode{{\rm V}_{cb}}\else{{\rm V}$_{cb}$\ }\fi}
\def\Vtb{\relax\ifmmode{{\rm V}_{tb}}\else{{\rm V}$_{tb}$\ }\fi}

\def\Dstrnostrp{\relax\ifmmode{{\parone D}^{(\star)+}}\else{${\parone D}^{(\star)+}$\ }\fi}
\def\Dstrnostrm{\relax\ifmmode{{\parone D}^{(\star)-}}\else{${\parone D}^{(\star)-}$\ }\fi}
\def\Dstrnostrz{\relax\ifmmode{{\parone D}^{(\star)0}}\else{${\parone D}^{(\star)0}$\ }\fi}

\def\jpsi{\relax\ifmmode{J/\psi}\else{$J/\psi$\ }\fi}

\def\gam{\relax\ifmmode{\gamma}\else{$\gamma$\ }\fi}

\def\W{\relax\ifmmode{{\parone W}}\else{{\parone W}}\fi}
\def\Wp{\relax\ifmmode{{\parone W}^+}\else{{\parone W}$^+$\ }\fi}
\def\Wm{\relax\ifmmode{{\parone W}^-}\else{{\parone W}$^-$\ }\fi}
\def\Wpm{\relax\ifmmode{{\parone W}^\pm}\else{{\parone W}$^\pm$\ }\fi}
\def\Wmp{\relax\ifmmode{{\parone W}^\mp}\else{{\parone W}$^\mp$\ }\fi}

\def\Z{\relax\ifmmode{{\parone Z}}\else{{\parone Z}}\fi}
\def\ZZ{\relax\ifmmode{{\parone Z}^0}\else{{\parone Z}$^0$\ }\fi}

\def\nub{\relax\ifmmode{\bar{\nu}}
	\else{$\bar{\nu}$\ }\fi}

\def\nue{\relax\ifmmode{\nu_e}\else{$\nu_e$\ }\fi}
\def\nueb{\relax\ifmmode{\bar{\nu}\tiny_e}
	\else{$\bar{\nu}\tiny_e$\ }\fi}

\def\el{\relax\ifmmode{e}\else{$e$\ }\fi}
\def\elp{\relax\ifmmode{e^+}\else{$e^+$\ }\fi}
\def\elm{\relax\ifmmode{e^-}\else{$e^-$\ }\fi}
\def\elpm{\relax\ifmmode{e^\pm}\else{$e^\pm$\ }\fi}
\def\elmp{\relax\ifmmode{e^\mp}\else{$e^\mp$\ }\fi}
\def\epem{\relax\ifmmode{e^+e^-}\else{$e^+e^-$\ }\fi}

\def\numu{\relax\ifmmode{\nu_\mu}\else{$\nu_\mu$\ }\fi}
\def\numub{\relax\ifmmode{\bar{\nu}\tiny_\mu}
	\else{$\bar{\nu}\tiny_\mu$\ }\fi}

\def\nutau{\relax\ifmmode{\nu_\tau}\else{$\nu_\tau$\ }\fi}
\def\nutaub{\relax\ifmmode{\bar{\nu}\tiny_\tau}
	\else{$\bar{\nu}\tiny_\tau$\ }\fi}

\def\taup{\relax\ifmmode{\tau^+}\else{$\tau^+$\ }\fi}
\def\taum{\relax\ifmmode{\tau^-}\else{$\tau^-$\ }\fi}
\def\taupm{\relax\ifmmode{\tau^\pm}\else{$\tau^\pm$\ }\fi}
\def\taump{\relax\ifmmode{\tau^\mp}\else{$\tau^\mp$\ }\fi}

\def\pip{\relax\ifmmode{\pi^+}\else{$\pi^+$\ }\fi}
\def\piz{\relax\ifmmode{\pi^0}\else{$\pi^0$\ }\fi}
\def\pizs{\relax\ifmmode{\pi^0\rm{s}}\else{$\pi^0\rm{s}$\ }\fi}
\def\pim{\relax\ifmmode{\pi^-}\else{$\pi^-$\ }\fi}
\def\pipm{\relax\ifmmode{\pi^\pm}\else{$\pi^\pm$\ }\fi}
\def\pimp{\relax\ifmmode{\pi^\mp}\else{$\pi^\mp$\ }\fi}
\def\pipmz{\relax\ifmmode{\pi^{\pm,0}}\else{$\pi^{\pm,0}$\ }\fi}

\def\etap{\relax\ifmmode{\eta^{\prime}}\else{$\eta^{\prime}$\ }\fi}

\def\rhop{\relax\ifmmode{\rho^+}\else{$\rho^+$\ }\fi}
\def\rhoz{\relax\ifmmode{\rho^0}\else{$\rho^0$\ }\fi}
\def\rhom{\relax\ifmmode{\rho^-}\else{$\rho^-$\ }\fi}
\def\rhopm{\relax\ifmmode{\rho^\pm}\else{$\rho^\pm$\ }\fi}
\def\rhomp{\relax\ifmmode{\rho^\mp}\else{$\rho^\mp$\ }\fi}
\def\rhopmz{\relax\ifmmode{\rho^{\pm,0}}\else{$\rho^{\pm,0}$\ }\fi}

\def\omegz{\relax\ifmmode{\omega^0}\else{$\omega^0$\ }\fi}

\def\aone{\relax\ifmmode{a_{1}}\else{$a_{1}$\ }\fi}
\def\aonep{\relax\ifmmode{a^{+}_{1}}\else{$a^{+}_{1}$\ }\fi}
\def\aonez{\relax\ifmmode{a^{0}_{1}}\else{$a^{0}_{1}$\ }\fi}
\def\aonem{\relax\ifmmode{a^{-}_{1}}\else{$a^{-}_{1}$\ }\fi}
\def\aonepm{\relax\ifmmode{a^{\pm}_{1}}\else{$a^{\pm}_{_1}$\ }\fi}
\def\aonemp{\relax\ifmmode{a^{\mp}_{1}}\else{$a^{\mp}_{1}$\ }\fi}

\def\K{\relax\ifmmode{{\parone K}}\else{{\parone K}}\fi}
\def\Kb{\relax\ifmmode{\bar{{\parone K}}}
	\else{$\bar{{\parone K}}$\ }\fi}

\def\Kz{\relax\ifmmode{{\parone K}^0}\else{{\parone K}$^0$\ }\fi}
\def\Ksh{\relax\ifmmode{{\parone K}^0_S}\else{{\parone K}$^0_S$\ }\fi}
\def\Klo{\relax\ifmmode{{\parone K}^0_L}\else{{\parone K}$^0_L$\ }\fi}
\def\Kzb{\relax\ifmmode{\bar{{\parone K}}\tiny^0}
	\else{$\bar{{\parone K}}\tiny^0$\ }\fi}

\def\Kpm{\relax\ifmmode{{\parone K}^\pm/{\parone \pi}^\pm}\else{${\parone K}^\pm/{\parone \pi}^\pm$\ }\fi}

\def\Kp{\relax\ifmmode{{\parone K}^+}\else{{\parone K}$^+$\ }\fi}
\def\Km{\relax\ifmmode{{\parone K}^-}\else{{\parone K}$^-$\ }\fi}
\def\Kpm{\relax\ifmmode{{\parone K}^\pm}\else{{\parone K}$^\pm$\ }\fi}
\def\Kmp{\relax\ifmmode{{\parone K}^\mp}\else{{\parone K}$^\mp$\ }\fi}

\def\D{\relax\ifmmode{{\parone D}}\else{${\parone D}$\ }\fi}

\def\Db{\relax\ifmmode{\bar{{\parone D}}}\else{$\bar{{\parone D}}$\ }\fi}

\def\Dz{\relax\ifmmode{{\parone D}^0}\else{{\parone D}$^0$\ }\fi}
\def\Dzb{\relax\ifmmode{\bar{{\parone D}}\tiny^0}
	\else{$\bar{{\parone D}}\tiny^0$\ }\fi}

\def\Dp{\relax\ifmmode{{\parone D}^+}\else{{\parone D}$^+$\ }\fi}
\def\Dm{\relax\ifmmode{{\parone D}^-}\else{{\parone D}$^-$\ }\fi}
\def\Dpm{\relax\ifmmode{{\parone D}^\pm}\else{{\parone D}$^\pm$\ }\fi}
\def\Dmp{\relax\ifmmode{{\parone D}^\mp}\else{{\parone D}$^\mp$\ }\fi}

\def\prtn{\relax\ifmmode{{\parone p}}\else{{\parone p}}\fi}
\def\prtnb{\relax\ifmmode{\bar{{\parone p}}}
	\else{$\bar{{\parone p}}$\ }\fi}

\def\ntrn{\relax\ifmmode{{\parone n}}\else{{\parone n}}\fi}
\def\ntrnb{\relax\ifmmode{\bar{{\parone n}}}
	\else{$\bar{{\parone n}}$\ }\fi}

\def\lam{\relax\ifmmode{\Lambda}\else{$\Lambda$\ }\fi}
\def\lamb{\relax\ifmmode{\bar{\Lambda}}
	\else{$\bar{\Lambda}$\ }\fi}
\def\lamz{\relax\ifmmode{\Lambda^0}\else{$\Lambda^0$\ }\fi}
\def\lamzb{\relax\ifmmode{\bar{\Lambda}\tiny^0}
	\else{$\bar{\Lambda}\tiny^0$\ }\fi}

\def\US{\relax\ifmmode{\Upsilon(1S)}
			\else{$\Upsilon(1S)$\ }\fi}
\def\USS{\relax\ifmmode{\Upsilon(2S)}
	\else{$\Upsilon(2S)$\ }\fi}
\def\USSS{\relax\ifmmode{\Upsilon(3S)}
	\else{$\Upsilon(3S)$\ }\fi}
\def\USSSS{\relax\ifmmode{\Upsilon(4S)}
	\else{$\Upsilon(4S)$\ }\fi}
\def\USSSSS{\relax\ifmmode{\Upsilon(5S)}
	\else{$\Upsilon(5S)$\ }\fi}

\def\Kstr{\relax\ifmmode{{\parone K}^\star}\else{{\parone K}$^\star$\ }\fi}
\def\Kstrb{\relax\ifmmode{\b
ar{{\parone K}}\tiny^\star}
	\else{$\bar{{\parone K}}\tiny^\star$\ }\fi}

\def\Kstrz{\relax\ifmmode{{\parone K}^{\star0}}\else{{\parone K}$^{\star0}$\ }\fi}
\def\Kstrzb{\relax\ifmmode{\bar{{\parone K}}\tiny^{\star0}}
	\else{$\bar{{\parone K}}\tiny^{\star0}$\ }\fi}

\def\Kstrp{\relax\ifmmode{{\parone K}^{\star+}}\else{{\parone K}$^{\star+}$\ }\fi}
\def\Kstrm{\relax\ifmmode{{\parone K}^{\star-}}\else{{\parone K}$^{\star-}$\ }\fi}
\def\Kstrpm{\relax\ifmmode{{\parone K}^{\star\pm}}\else{{\parone K}$^{\star\pm}$\ }\fi}
\def\Kstrmp{\relax\ifmmode{{\parone K}^{\star\mp}}\else{{\parone K}$^{\star\mp}$\ }\fi}

\def\Dstr{\relax\ifmmode{{\parone D}^\star}\else{{\parone D}$^\star$\ }\fi}
\def\Dstrb{\relax\ifmmode{\bar{{\parone D}}\tiny^\star}
	\else{$\bar{{\parone D}}\tiny^\star$\ }\fi}

\def\Dstrz{\relax\ifmmode{{\parone D}^{\star 0}}\else{{\parone D}$^{\star 0}$\
}\fi}

\def\Dstrzb{\relax\ifmmode{\bar{{\parone D}}\tiny^{\star 0}}
	\else{$\bar{{\parone D}}\tiny^{\star 0}$\ }\fi}

\def\Dstrp{\relax\ifmmode{{\parone D}^{\star+}}\else{{\parone D}$^{\star+}$\ }\fi}
\def\Dstrm{\relax\ifmmode{{\parone D}^{\star-}}\else{{\parone D}$^{\star-}$\ }\fi}
\def\Dstrpm{\relax\ifmmode{{\parone D}^{\star\pm}}\else{{\parone D}$^{\star\pm}$\ }\fi}
\def\Dstrmp{\relax\ifmmode{{\parone D}^{\star\mp}}\else{{\parone D}$^{\star\mp}$\ }\fi}

\def\Ddblowz{\relax\ifmmode{D_1(2420)^0}\else{$D_1(2420)^0$\ }\fi}
\def\Ddbhighz{\relax\ifmmode{D_2^*(2460)^0}\else{$D^*_2(2460)^0$\ }\fi}
\def\Ddblow{\relax\ifmmode{D_1(2420)}\else{$D_1(2420)$\ }\fi}
\def\Ddbhigh{\relax\ifmmode{D_2^*(2460)}\else{$D^*_2(2460)$\ }\fi}
\def\Ddblowp{\relax\ifmmode{D_1(2420)^+}\else{$D_1(2420)^+$\ }\fi}
\def\Ddbhighp{\relax\ifmmode{D_2^*(2460)^+}\else{$D^*_2(2460)^+$\ }\fi}
\def\Ddblowm{\relax\ifmmode{D_1(2420)^-}\else{$D_1(2420)^-$\ }\fi}
\def\Ddbhighm{\relax\ifmmode{D_2^*(2460)^-}\else{$D^*_2(2460)^-$\ }\fi}
\def\DJ{\relax\ifmmode{D_J}\else{$D_J$\ }\fi}

\def\B{\relax\ifmmode{\parone{B}}\else{$\parone{B}$\ }\fi}
\def\Bb{\relax\ifmmode{\bar{\parone{B}}}\else{$\bar{{\parone B}}$\ }\fi}

\def\Bz{\relax\ifmmode{{\parone B}^0}\else{${\parone B}^0$\ }\fi}
\def\Bzb{\relax\ifmmode{\bar{\parone B}^0}\else{$\bar{\parone B}^0$\ }\fi}

\def\Bp{\relax\ifmmode{\parone{B}^+}\else{$\parone{B}^+$\ }\fi}
\def\Bm{\relax\ifmmode{\parone{B}^-}\else{$\parone {B}^-$\ }\fi}

\def\bbar{\relax\ifmmode{B\bar{B}}\else{$B\bar{B}$\ }\fi}
\def\qqbar{\relax\ifmmode{q\bar{q}}\else{$q\bar{q}$\ }\fi}

\def\dzpiz{\relax\ifmmode{\Dstrz \to \Dz \piz}\else
	{$\Dstrz \to \Dz \piz$\ }\fi}
\def\dzpip{\relax\ifmmode{D^{*+} \to D^0\pi^+}\else 
	{$D^{*+} \to D^0\pi^+$\ }\fi}

\def\dzgam{\relax\ifmmode{\Dstrz \to \Dz \gamma}\else
	{$\Dstrz \to \Dz \gamma$\ }\fi}

\def\Kmpip{\relax\ifmmode{\D^0 \to \Km \pip}\else
	{$\D^0 \to \Km \pip$\ }\fi}
\def\Kmpippiz{\relax\ifmmode{\D^0 \to \Km \pip \piz}\else
	{$\D^0 \to \Km \pip \piz$\ }\fi}
\def\Kmpippimpip{\relax\ifmmode{\D^0 \to \Km \pip \pim \pip}\else
	{$\D^0 \to \Km \pip \pim \pip$\ }\fi}
\def\Kmpippip{\relax\ifmmode{\Dp \to \Km \pip \pip}\else
	{$\Dp \to \Km \pip \pip$\ }\fi}

\def\Kpi{\relax\ifmmode{\D^0 \to \Km \pip}\else
	{$\D^0 \to \Km \pip$\ }\fi}
\def\Kpipiz{\relax\ifmmode{\D^0 \to \Km \pip \piz}\else
	{$\D^0 \to \Km \pip \piz$\ }\fi}
\def\Kpipipi{\relax\ifmmode{\D^0 \to \Km \pip \pim \pip}\else
	{$\D^0 \to \Km \pip \pim \pip$\ }\fi}
\def\Kpipi{\relax\ifmmode{\Dp \to \Km \pip \pip}\else
	{$\Dp \to \Km \pip \pip$\ }\fi}

\def\decrhop{\relax\ifmmode{\rhop \to \pip \piz}\else
	{$\rhop \to \pip \piz$\ }\fi}
\def\decomega{\relax\ifmmode{\omega \to \pip \pim \piz}\else
	{$\omega \to \pip \pim \piz$\ }\fi}
\def\deceta{\relax\ifmmode{\eta \to \gamm \gamm}\else
	{$\eta \to \gamm \gamm$\ }\fi}
\def\decetap{\relax\ifmmode{\etap \to \eta \gamm \gamm}\else
	{$\etap \to \eta \gamm \gamm$\ }\fi}

\def\Dpi{\relax\ifmmode{\Bzb \to \Dp \pim}\else
	{$\Bzb \to \Dp \pim$\ }\fi}
\def\Drho{\relax\ifmmode{\Bzb \to \Dp \rhom}\else
	{$\Bzb \to \Dp \rhom$\ }\fi}
\def\Daone{\relax\ifmmode{\Bzb \to \Dp \aonem}\else
    	{$\Bzb \to \Dp \aonem$\ }\fi}

\def\Dstrpi{\relax\ifmmode{\Bzb \to \Dstrp \pim}\else
	{$\Bzb \to \Dstrp \pim$\ }\fi}
\def\Dstrrho{\relax\ifmmode{\Bzb \to \Dstrp\rhom}\else
	{$\Bzb \to \Dstrp\rhom$\ }\fi}
\def\Dstraone{\relax\ifmmode{\Bzb \to \Dstrp \aonem}\else
	{$\Bzb \to \Dstrp \aonem$\ }\fi}

\def\Dzpiz{\relax\ifmmode{\Bzb \to \Dz \piz}\else
	{$\Bzb \to \Dz \piz$\ }\fi}
\def\Dzrhoz{\relax\ifmmode{\Bzb \to \Dz \rhoz}\else
	{$\Bzb \to \Dz \rhoz$\ }\fi}
\def\Dzeta{\relax\ifmmode{\Bzb \to \Dz \eta}\else
	{$\Bzb \to \Dz \eta$\ }\fi}
\def\Dzetap{\relax\ifmmode{\Bzb \to \Dz \etap}\else
	{$\Bzb \to \Dz \etap$\ }\fi}
\def\Dzomega{\relax\ifmmode{\Bzb \to \Dz \omega}\else
	{$\Bzb \to \Dz \omega$\ }\fi}
\def\Dzaonez{\relax\ifmmode{\Bzb \to \Dz \aonez}\else
	{$\Bzb \to \Dz \aonez$\ }\fi}

\def\Dstrzpiz{\relax\ifmmode{\Bzb \to \Dstrz \piz}\else
	{$\Bzb \to \Dstrz \piz$\ }\fi}
\def\Dstrzrhoz{\relax\ifmmode{\Bzb \to \Dstrz \rhoz}\else
	{$\Bzb \to \Dstrz \rhoz$\ }\fi}
\def\Dstrzeta{\relax\ifmmode{\Bzb \to \Dstrz \eta}\else
	{$\Bzb \to \Dstrz \eta$\ }\fi}
\def\Dstrzetap{\relax\ifmmode{\Bzb \to \Dstrz \etap}\else
	{$\Bzb \to \Dstrz \etap$\ }\fi}
\def\Dstrzomega{\relax\ifmmode{\Bzb \to \Dstrz \omega}\else
	{$\Bzb \to \Dstrz \omega$\ }\fi}
\def\Dstrzaonez{\relax\ifmmode{\Bzb \to \Dstrz \aonez}\else
	{$\Bzb \to \Dstrz \aonez$\ }\fi}

\def\Dzpi{\relax\ifmmode{\Bm \to \Dz \pim}\else
	{$\Bm \to \Dz \pim$\ }\fi}
\def\Dzrho{\relax\ifmmode{\Bm \to \Dz \rhom}\else
	{$\Bm \to \Dz \rhom$\ }\fi}
\def\Dzaone{\relax\ifmmode{\Bm \to \Dz \aonem}\else
	{$\Bm \to \Dz \aonem$\ }\fi}

\def\Dstrzpi{\relax\ifmmode{\Bm \to \Dstrz \pim}\else
	{$\Bm \to \Dstrz \pim$\ }\fi}
\def\Dstrzrho{\relax\ifmmode{\Bm \to \Dstrz \rhom}\else
	{$\Bm \to \Dstrz \rhom$\ }\fi}
\def\Dstrzaone{\relax\ifmmode{\Bm \to \Dstrz \aonem}\else
	{$\Bm \to \Dstrz \aonem$\ }\fi}

\def\Dstrzgpi{\relax\ifmmode{\Bm \to \Dstrz \pim}\else
	{$\Bm \to \Dstrz \pim$\ }\fi}
\def\Dstrzrho{\relax\ifmmode{\Bm \to \Dstrz \rhom}\else
	{$\Bm \to \Dstrz \rhom$\ }\fi}

\def\PPNP#1#2#3{{\em Prog. Part. Nucl. Phys.} {\bf #1}, #2 (#3)}

\def\PRL#1#2#3{{\em Phys. Rev. Lett.} {\bf #1}, #2 (#3)}
\def\PRD#1#2#3{{{\em Phys. Rev.} D} {\bf #1}, #2 (#3)}
\def\PLB#1#2#3{{{\em Phys. Lett.}  B} {\bf #1}, #2 (#3)}

\def\mco{\multicolumn}

\def\be{\begin{equation}}
\def\ee{\end{equation}}
\def\bea{\begin{eqnarray}}
\def\eea{\end{eqnarray}}

\def\fig{Fig.}
\def\tbl{Table}

\begin{document}
\title{$b \to c$ HADRONIC DECAYS}
\author{ JORGE L. RODRIGUEZ}
\address{Department of Physics and Astronomy, University of Hawaii, 2505
Correa Road,\\ Honolulu, Hawaii 96822, USA \\ University of Hawaii preprint UH 511-892-98}
\maketitle\abstracts{
A review of current experimental results on exclusive hadronic 
decays of bottom mesons to a single or double charmed final state is
presented. We concentrate on branching fraction measurements conducted
at $e^+e^-$ colliders at the \USSSS and at the $Z^0$ resonance. The
experimental results reported are then used in tests of theoretical
model predictions, the determination of the QCD parameters $a_1$ and
$a_2/a_1$ and tests of factorization.}

\section{Introduction}

\begin{figure}[t]
\centerline{\epsfxsize 10.0 truecm\epsfbox{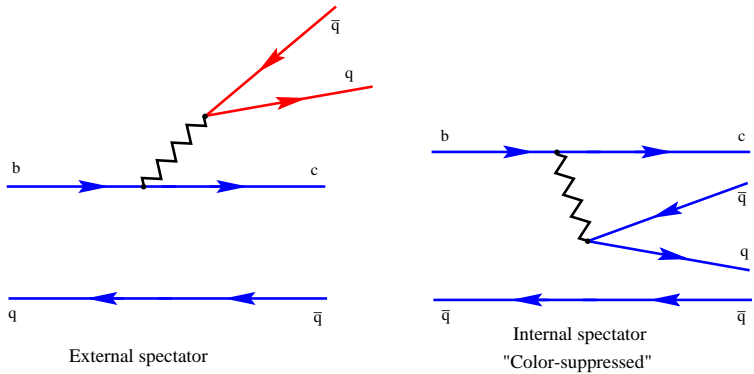}}
\caption{\label{fynfig1} Tree level decay diagrams dominant in two-body
hadronic decays of $B$ mesons.}
\end{figure}

In $b\to c$ hadronic hadronic transitions the spectator processes
dominate. Other processes, such as the exchange or annihilation
channels are suppressed relative to the spectator processes through
form-factor suppression~\cite{NS97}. Two classes of spectator processes
are possible --internal and external spectator (See \fig~\ref{fynfig1})--
each defined by the quark-color arrangement of the final state. In the
factorization approximation three classes of spectator decays can be
identified. In Class I or Class II decays, the quark configuration of
the final states, neglecting final-state rescattering, is possible only
if the decay proceeds either through the external or internal diagram
respectively. A third type of decay, Class III, the quark configuration
of the final-states can be obtained via both diagrams. 

In this article I will describe the latest experimental results on the
following hadronic two-body decays of the \B meson:
$$
\begin{array}{lll}
\mco{1}{c}{\rm Class\ I}& \mco{1}{c}{\rm Class\ II}&\mco{1}{c}{\rm Class\ III}\\
 B^0 \to D^{(*)+}(n\pi)^-	& B^0 \to D^{(*)0}(n\pi)^0	&B^- \to D^{(*)0}(n\pi)^-	\\
 B^0 \to D_J^+(n\pi)^-		& 				&B^- \to D_J^0(n\pi)^-	\\    
 B^0 \to D^{(*)+}D_s^{(*)-}	&				&B^- \to D^0K^-		\\
\end{array}
$$
concentrating on the experimental techniques employed in
measurements of branching fractions. I will also discuss some of the
more important theoretical implications including tests of factorization
and determination of the QCD parameters $a_1$ and $a_2/a_1$.

\section{Experimental Programs}
Except for the $B\to J/\psi K $ decays, measurements of hadronic decay
rates of the $B$ meson to charm have been performed exclusively in $e^+e^-$
experiments. Decay rate measurements are dominated by experiments
conducted at \bbar threshold while experiments at higher energy have
provide information on $b$ lifetimes and on higher mass hadrons e.g.,
 $B_s$, $\Lambda_b$ not accessible to experiments at \bbar 
threshold. Recently, experiments at CERN have made contributions to
measurement of exclusive decays with a measurements from OPAL on \Dstrpi
and ALEPH measurements of several $B\to DD_sX,DDX$ and two-body
$D^{(*)}D_s^{(*)}$ decays. In the next section I will briefly
describe the OPAL measurement as an example of the experimental method
employed at the high energy \epem experiments. I will then
concentrate on recent measurements performed at \cleo since they 
are the dominant contributors to measurements of exclusive hadronic 
decays of $B_{d,u}$ mesons.

\subsection{Measurement of the branching fraction of \Dstrpi at OPAL} 
The large sample of $Z^0$ produced at LEP coupled with a reasonable
partial width to $b\bar{b}$ (6\%) provides a significant number of \bbar 
events to examine. Unfortunately the small branching fractions to any
particular \B decay mode and the large particle multiplicities involved
present difficult obstacles in exclusive reconstruction of hadronic \B
decays. Given these obstacles OPAL has a new measurement of \Dstrpi with
their sample of $1.2 \times 10^6$ hadronic $Z^0$ decays. 

The analysis~\cite{OPAL} features a full reconstruction of the decay
chain:~\footnote{Charge-conjugate states are implied throughout this
article}
$$
B^0\to D^{*+}\pi^-;\ \ D^{*+}\to \Dz\pi^+;\ \ D^0\to \Km\pip
$$
where charged tracks are combined to form candidate particles.  Each
track in the event is identified as either a kaon or pion depending
primarily on its energy loss in the jet chamber. Additional requirements
on track transverse momentum and impact parameter are imposed. The
invariant mass of the $K^-\pi^+$ combination is required to be within 90
MeV of the nominal \Dz mass and the mass difference between the $K^-\pi^+$
and $(\Km\pip)\pip$ combination is required to be within 2 \mMeV of the
known mass difference. The \Dz and \Dstrp candidates are formed
separately for each jet and are then combined with other tracks in the
jet to form a $B_d$ candidate. To reduce combinatorial backgrounds, a
decay angle cut is imposed on the \Dz candidates and a helicity angle
cut is imposed on the angular distribution of the decay products of the
$\Dstrp.$

\begin{figure}
\vskip -.8cm
\centerline{\epsfxsize 7.0 truecm\epsfbox{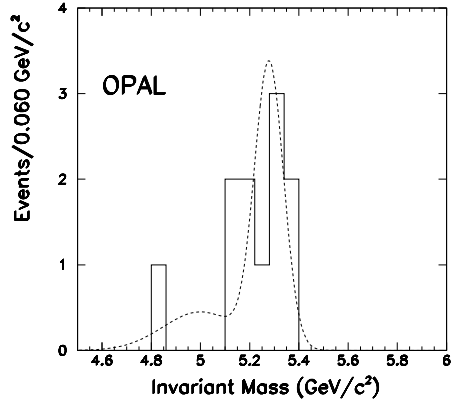}}
\vskip -1.cm
\caption{\label{fig_opal} The invariant mass distribution of fully
reconstructed $B_d$ mesons at OPAL. The dashed curve is the fit to data. 
It is a sum of two Gaussians plus a straight line. The first Gaussian
represents the signal the other the feed-down from the \Dstrrho decays. 
The straight line models the combinatoric background.}
\end{figure}

The hard fragmentation of the $b$ quark and its long lifetime are
also exploited to reduce combinatorial background. The momentum of the
$B_d$ candidate is required to exceed 70\% of the beam energy and the 
$B_d$ and \Dz decay vertices are required to be in the hemisphere centered
around the $B_d$ momentum vector.

After application of the selection criteria, 11 events are observed in
the mass region between 4.5 and 6.0 $\mGeV,$ see \fig~\ref{fig_opal}. To
determine the event yield the $m_B$ distribution is fit to two-Gaussians
plus a straight line. The second Gaussian takes into account the
feed-down from \Dstrrho decays which have been miss-reconstructed as a
 $\Dstrpi.$ This background peaks below the $B_d$ mass and is fixed in the
fit to the value determined in a Monte Carlo simulation. Monte Carlo data is
also used to determine the widths of both Gaussians. The fit yields
$8.1\pm 2.9$ events for the \Dstrpi and $2.9 \pm 1.9$ for the \Dstrrho
mode at a mass of $5.279 \pm 0.023\ \mGeV.$ The branching fraction is
estimated to be: $$ 
{\cal B}r(\Dstrpi) = (1.0 \pm 0.4 \pm 0.1) \%
$$
The Standard Model value for $\Gamma_{bb}/\Gamma_{\rm had}= 0.217$
and the $B_d$ production fraction $(f_{B_d} = 0.38)$ are used~\cite{OPAL}. 
This value is consistent with previous CLEO and ARGUS
results~\cite{PDG96} and with current \cleo results~\cite{JRconf}.

\section{Measurements at \bbar Threshold}
The large luminosity and clean environment typical at \epem machines
running at the \bbar threshold provide very large samples of \bbar events. For
example, at CLEO the numbers of \bbar events collected from 1994 through
1996 was $3.1 \times 10^6$ while ARGUS whose physics runs ended in 1993,
collected about 330,000 \bbar pairs. 

The event topology at \bbar threshold is somewhat different than at
a machine running on the $Z^0$ resonance. In particular, two
characteristics differentiate the reconstruction technique from that
employed in the higher energy environment. First, significantly larger
backgrounds are encountered from continuum processes. At \bbar threshold
3/4 of the total hadronic cross-section is from continuum
events. Fortunately, this background is well behaved and can be
accurately modeled by data taken below \bbar threshold; nevertheless,
techniques must be employed to reduce the contribution from these
backgrounds. Secondly, the fact that the \B mesons are produced nearly at
rest is exploited to improve signal identification. Signal extraction is
improved by replacing the energy of the reconstructed \B meson with the
energy of the beam which is typically known better by an order of
magnitude than the reconstructed energy.

In the following sections I will discuss new results from CLEO using the
complete \cleo data sample which consists of $3.1 fb^{-1}$ taken at the 
 $\USSSS.$ A smaller sample of $1.4 fb^{-1}$ taken just below the \USSSS is
also used to model the continuum background. These results have either been
reported at conferences or have been recently published. All of the
\B branching fractions presented in Tables \ref{tbl_dnpi},\ref{tbl_color} and
\ref{tbl_DDs} have been rescaled to the $\Dz,\Dp,\Dstrp$ and \Dstrz 
branching fractions used in Ref. [8]. The $D_s^{(*)}$ branching
fractions used are taken from Ref. [9].  

\begin{figure}[t]
\vskip -1.cm
\centerline{\psfig{figure=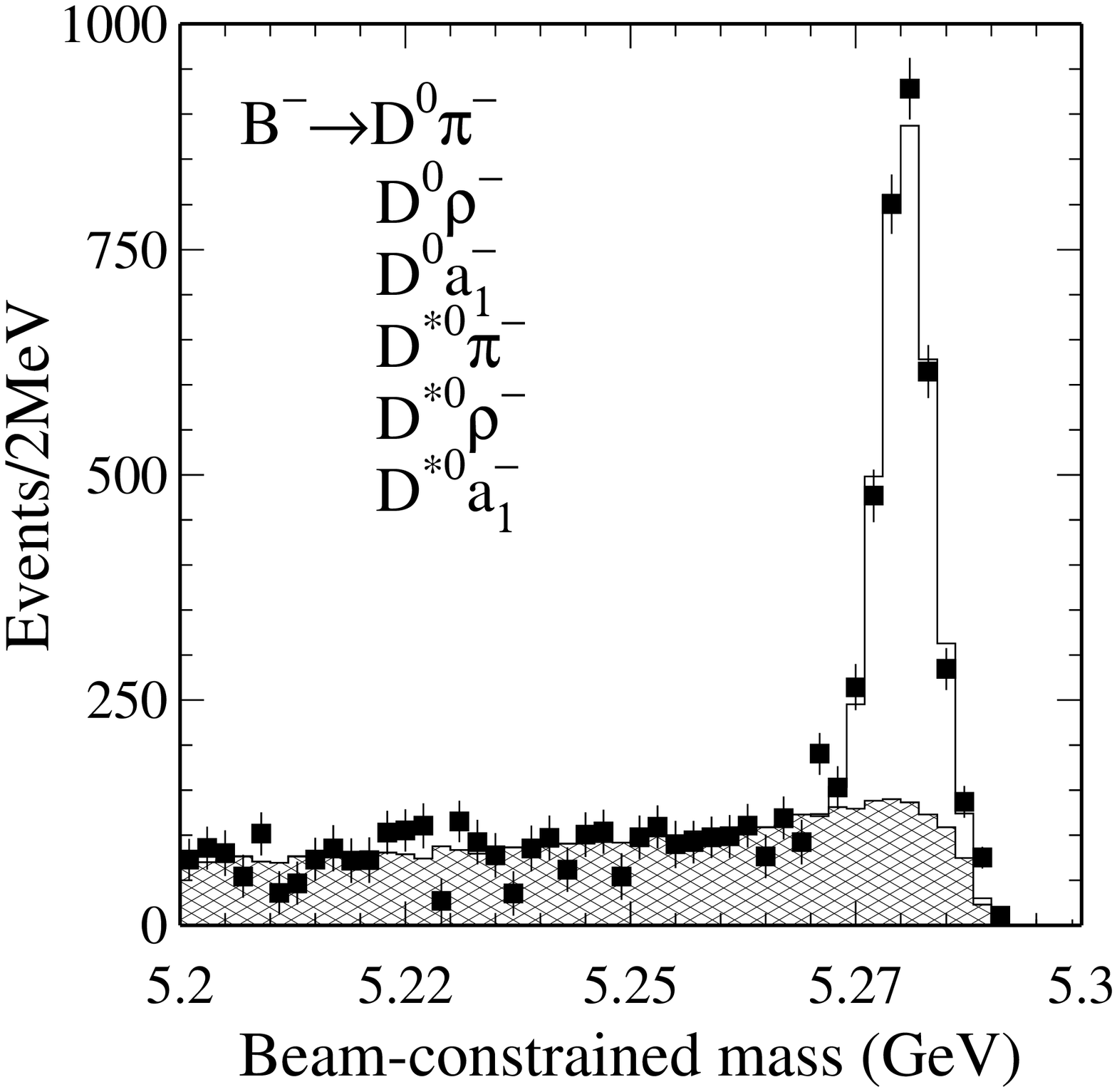,width=5.0cm}\psfig{figure=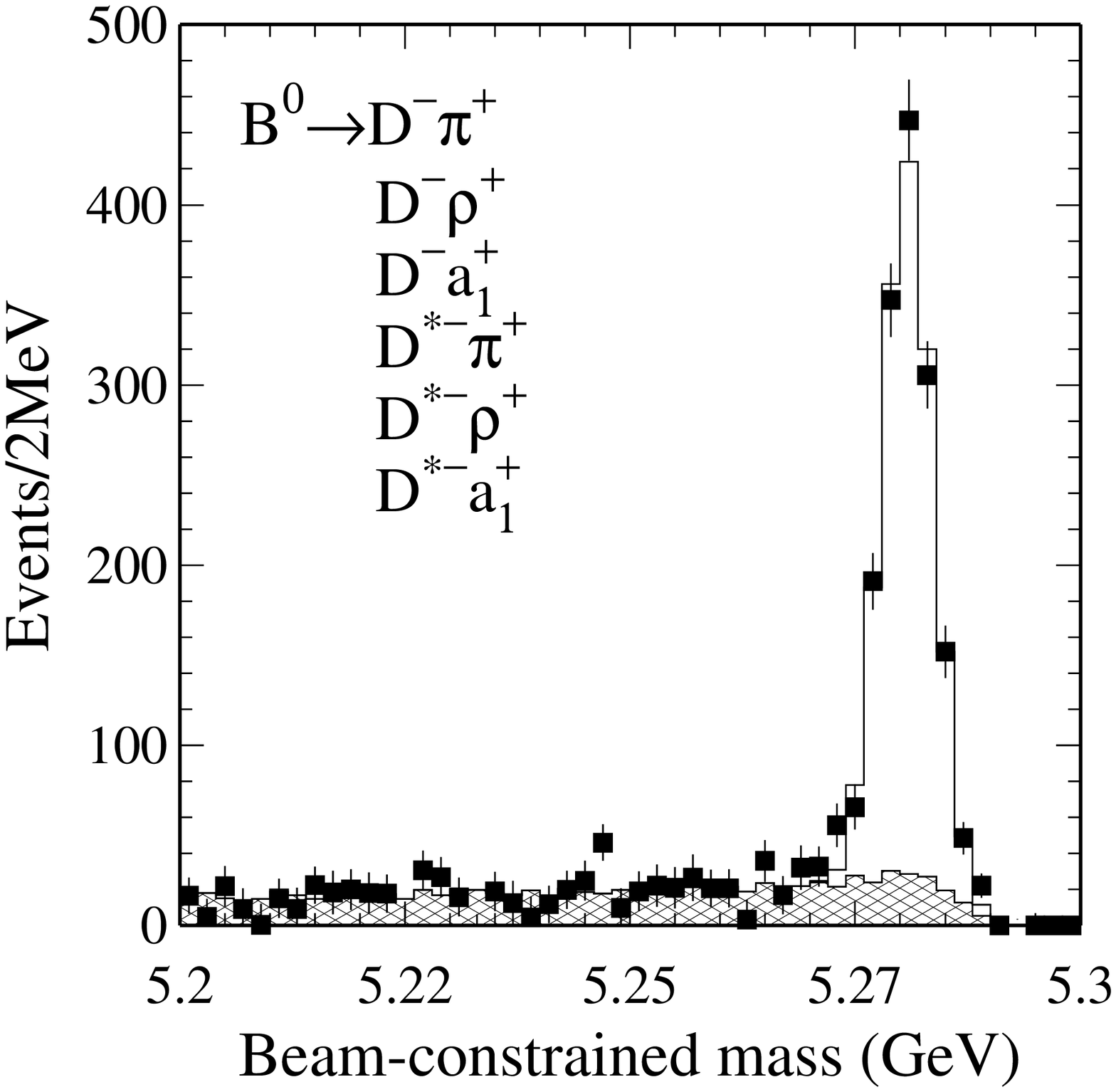,width=5.0cm}}
\caption{\label{fig_dnpi}The continuum-subtracted beam-constrained mass
distributions for twelve $B \to D^{(*)}n\pi$ modes. The hatched
histograms shows the \bbar background spectrum, the open histograms is
fit to data by the sum of a Gaussian of plus the \bbar background
distribution determined from Monte Carlo simulations. The data are the
black squares with error bars.}
\end{figure}

\subsection{Full Reconstruction of $B \to D^{(*)}(n\pi)$ decays at CLEO}
CLEO has recently updated their branching fraction measurements of 22 decay
modes of the $B_{d,u}$ mesons. These have been released as conference
reports~\cite{JRconf,colorconf}.  The analyses described here utilize the
full reconstruction technique, similar to the method described in the
previous section but optimized to exploit the kinematic properties
unique to \bbar production from the decay of the \USSSS. Once again the
goal is to maximize the number of \B mesons by combining a selection of
charged and neutral tracks into particle candidates which in turn are
combined to reconstruct the \B meson decay chain. The charmed
candidates are formed from a selected sample of charged tracks and \piz
in the following decay modes: 
$$
\begin{array}{ll}
D^{*0} \to D^0\pi^0 & D^{*+} \to D^0\pi^+ \\
D^0 \to K^-\pi^+, K^-\pi^+\pi^0, K^-\pi^+\pi^-\pi^+ & D^+\to
K^-\pi^+\pi^+.
\end{array}
$$
Invariant mass cuts are applied to the \Dz and \Dm candidates and the
$\Dstr,$\ ~\Dz mass difference is used to select \Dstrp and \Dstrz candidates.

To reduce continuum background two quantities are used. The global event
topology is exploited by cutting on the ratio of the $2^{nd}$ to
 $1^{st}$ Fox-Wolfram moment, selecting events above 0.5. Second, the
sphericity angle $\theta_f$ is used. It too exploits the difference
between the event topology of continuum and \bbar events. The sphericity
angle is defined as the angle between the sphericity axis for tracks
which form the \B candidate and the sphericity axis for the remaining
tracks in the event. For real \B events this angle is isotropic thus the 
 $|\cos\theta_f|$ quantity is distributed uniformly from 0 to 1. For
continuum events this quantity populates the region near 1.0. By
selecting candidates with $|\cos\theta_f| \le 0.8$ a considerable
portion of the continuum background is rejected while losing only 20\%
of the signal.

\begin{table}
\begin{center}
\caption{\label{tbl_dnpi}Branching Fractions for $B\to D^{(*)}(n\pi)^-$ Decay Modes~\protect\cite{changes}} 
\begin{tabular}{|ll||ll|}\hline
 \mco{2}{|c||}{Class I Decays} &  \mco{2}{c|}{Class III Decays} \\
  Mode 	       	&\mco{1}{c||}{${\cal B}r$ (\%)}    & Mode 	&\mco{1}{c|}{${\cal B}r$ (\%)}\\ \hline
 $\Dp\pim$		   & $0.25\pm 0.02\pm 0.03$&$\Dz\pim$                 & $0.47\pm 0.03\pm 0.04$\\ 
 $\Dstrp\pim$~\cite{pfave} & $0.27\pm 0.01\pm 0.02$&$\Dstrz\pim$~\cite{pfave} & $0.40\pm 0.03\pm 0.04$\\
 $\Dp\rhom$~\cite{nrrho}   & $0.79\pm 0.07\pm 0.12$&$\Dz\rhom$~\cite{nrrho}   & $0.92\pm 0.08\pm 0.08$\\ 
 $\Dstrp\rhom$~\cite{nrrho}& $0.73\pm 0.06\pm 0.08$&$\Dstrz\rhom$~\cite{nrrho}& $1.28\pm 0.13\pm 0.14$\\ 
 $\Dp\aonem$~\cite{nra1}   & $0.83\pm 0.09\pm 0.14$&$\Dz\aonem$~\cite{nra1}   & $0.89\pm 0.10\pm 0.11$\\
 $\Dstrp\aonem$~\cite{nra1}& $1.16\pm 0.12\pm 0.16$&$\Dstrz\aonem$~\cite{nra1}& $1.60\pm 0.25\pm 0.24$\\
                           &                       &$D_1^0\pim$               & $0.12\pm 0.02\pm 0.03$\\
                           &                       &$D_2^{*0}\pim$            & $0.21\pm 0.08\pm 0.03$\\
                           &                       &$D^{0}K^-$  	      & $<0.044$              \\\hline
\end{tabular}					      
\end{center}
\end{table}					      
 
To reduce contamination from other \bbar decays the quantity \ediff
-- defined as the difference between the reconstructed energy and the
beam energy -- is required to lie within $\pm 2.5\sigma_{\Delta E}$ from
0.0. Since $\sigma_{\Delta E}$ is less than a pion
mass, -- 15 to 45 \mMeV -- this cut is very effective in reducing
the background contribution from misreconstructed \B decays which
differ by one or more pions.

The number of fully reconstructed \B in each mode is determined by
fitting the beam-constrained mass to a single Gaussian plus a background
shape. The beam-constrained mass variable is defined as the usual
invariant mass but with the energy of the \B replace by the beam
energy. The beam-constrained mass distributions for 12 decay modes,
separated into \Bz and \Bm modes, are shown in \fig~\ref{fig_dnpi}. These
decay modes are color-allowed, that is, they can proceed by the external
spectator diagram. The measured branching fractions are listed in
\tbl~\ref{tbl_dnpi}.

\cleo has also searched for color-suppressed decays
to a single charmed meson, $B^0\to D^{(*)0}X^0$. So far no clear signal
has been observed. The analysis presented in Ref. [10] uses
the full reconstruction technique but the signal extraction is done
using the \ediff variable instead of the beam-constrained mass
technique. The 90\% C.L. upper limits are listed in
\tbl~\ref{tbl_color}. They are updated results obtained with the complete
\cleo data sample and replace previous \cleo numbers.

\begin{table}
\begin{center}
\caption{\label{tbl_color}Limits on Branching Fractions for $\bar{B}^0\to D^{(*)0}(n\pi)^0$ Decay Modes~\protect\cite{changes}} 
\begin{tabular}{|ll||ll|}\hline
  \mco{4}{|c|}{Class II Decays @ 90\% C.L.} \\
  Mode &\mco{1}{c||}{${\cal B}r$ (\%)}	   & Mode 	&\mco{1}{c|}{${\cal B}r$ (\%)}\\ \hline
 $\Dz\piz$     &$<0.01$ 	& $\Dstrz\piz$  &$<0.04$  \\        
 $\Dz\eta$     &$<0.01$ 	& $\Dstrz\eta$  &$<0.02$  \\       
 $\Dz\etap$    &$<0.10$ 	& $\Dstrz\etap$ &$<0.16$  \\        
 $\Dz\rhoz$    &$<0.04$ 	& $\Dstrz\rhoz$ &$<0.06$  \\        
 $\Dz\omega$   &$<0.05$ 	& $\Dstrz\omega$&$<0.08$  \\ \hline
\end{tabular}					      
\end{center}
\end{table}

\subsection{Partial Reconstruction of $B \to D^{(*)}\pi^-$ decays at CLEO}
\label{Partial}

\cleo has recently measured the \Dstrpi and \Dstrzpi branching fraction
using the partial reconstruction technique~\cite{dstrpi-partial}.  In
this method the kinematics of \bbar production at the \USSSS are
exploited to fully reconstruct the decay using only the 4-momentum of
the fast pion ($\pi_f$) from the decay of the \B meson and the slow pion
($\pi_s$) from the \Dstr decay. The advantage of this method lies in the
large increase in statistical power since the explicit reconstruction of
the \D meson is no longer needed. The trade-off comes in the larger
backgrounds and hence greater difficulty in signal extraction.

The two pion 4-momenta together with inputs of the known masses, the $\D,$
\Dstr and \B meson masses, and requiring that the energy of the \B be
equal to the beam energy provide sufficient constraints to solve for the
unknown \Dstr and \Dz momentum.  Also, energy-momentum conservation in
the decay of the \B meson restricts the \Dstr momentum vector to lie in
a cone around the $\pi_f$. Similarly, for the \Dstr decay, the \Dstr
momentum vector must lie in a cone around the $\pi_s$. For the two pions
to be consistent with $B\to D^*\pi$ decay the cones must overlap. This
condition is required as part of the event selection criteria.

\begin{figure}[t]
\vskip -.75cm
\centerline{\psfig{figure=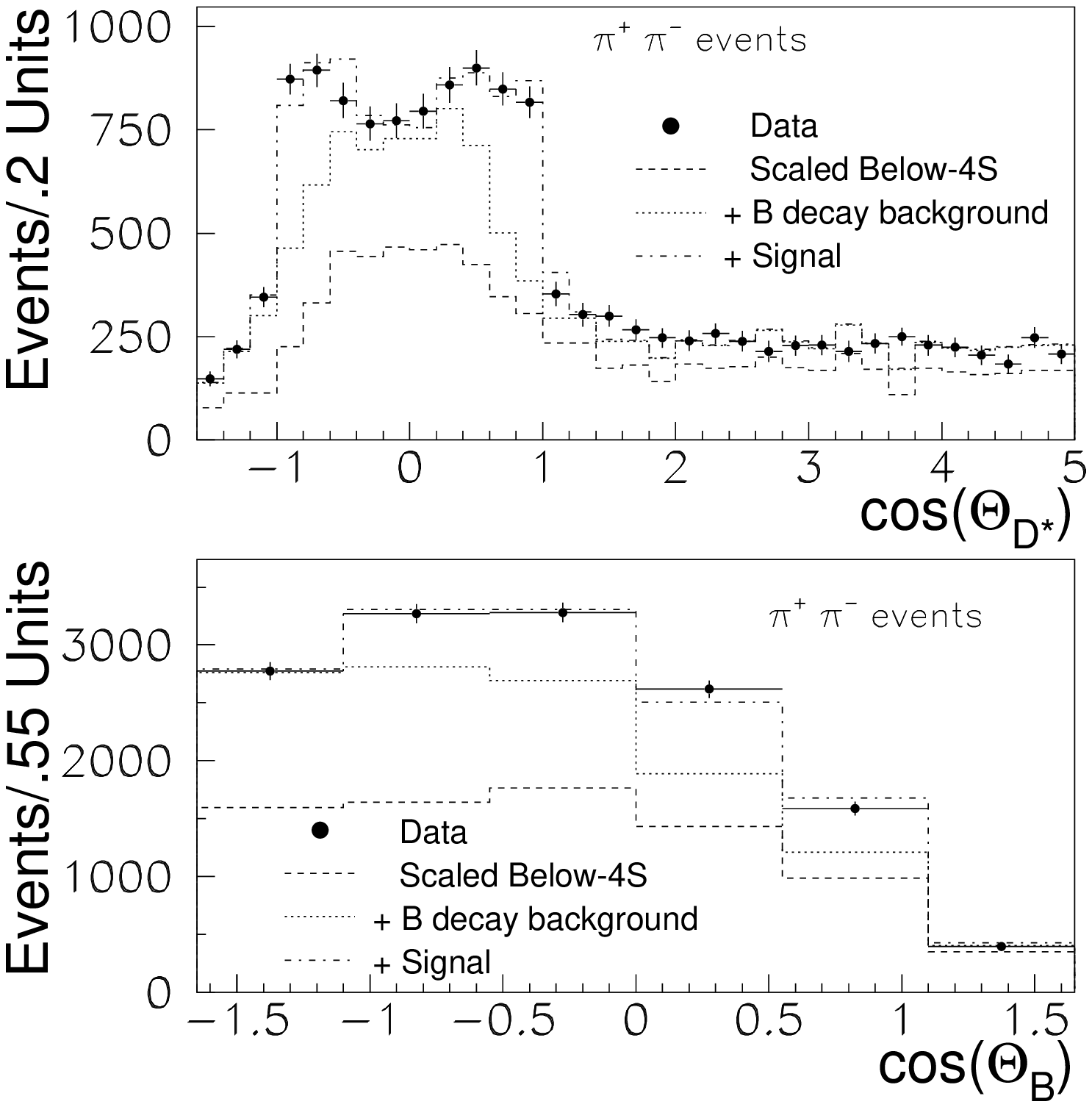,width=2.75in}\hspace{-.4in}\psfig{figure=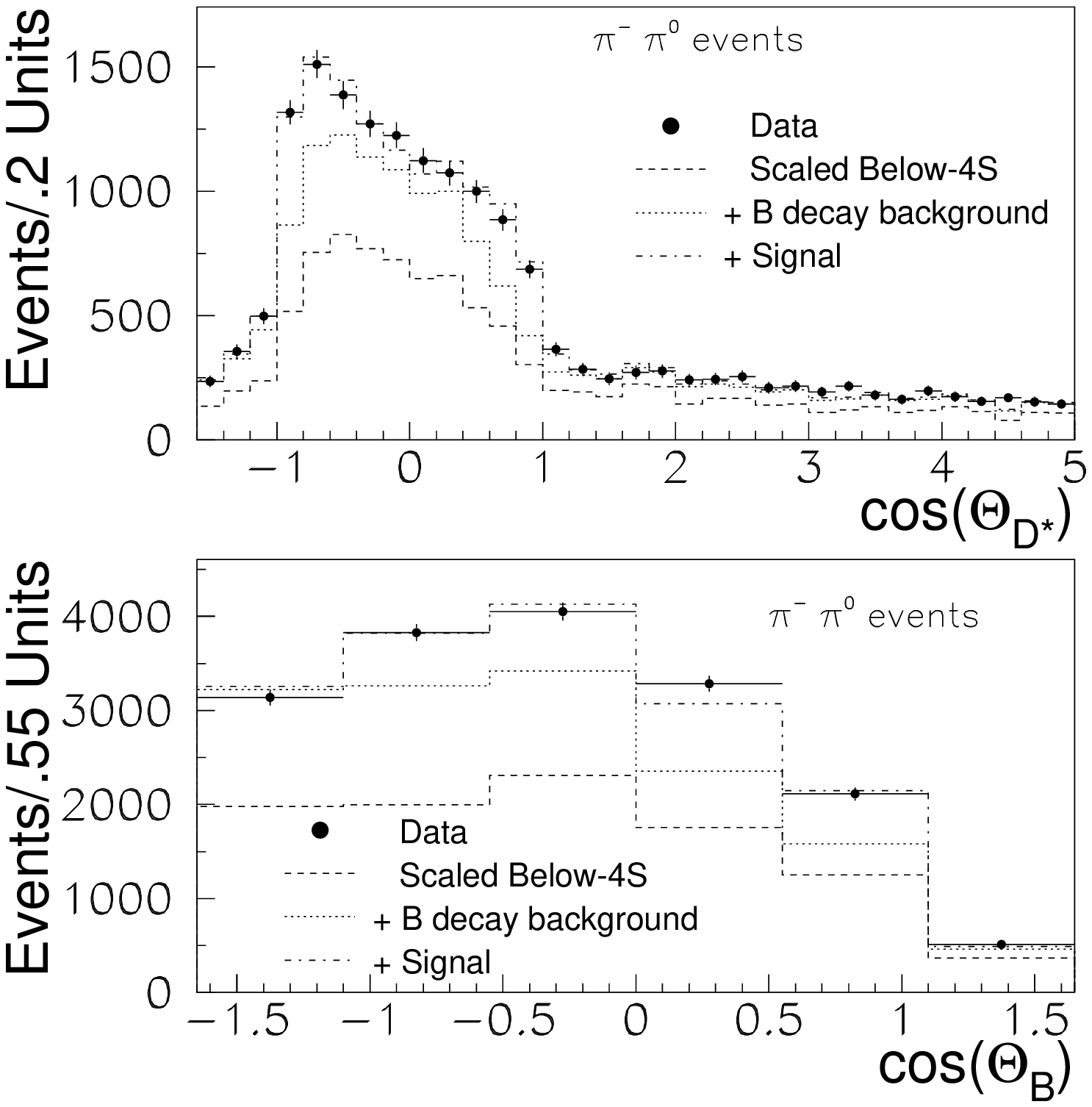,width=2.75in}}
\vskip -.8cm
\hspace{2.cm}$\Dstrpi$\hspace{1.6in}$\Dstrzpi$
\caption{\label{fig_dstrpi-part} Projections of 2-D fit to data for the
variables $\cos(\Theta_B)$ and $\cos(\Theta_B)$. The two figures on the left
represent \Dstrpi decays the two on the right are for \Dstrzpi
decays. The population of events in the non-physical region
$|\cos(\Theta)|> 1.0$ is from mismeasured track combinations.}
\end{figure}

Signal extraction in this analysis is performed by inspecting the
distributions of the $\cos(\Theta_{D^*})$ and $\cos(\Theta_{B})$ where
$\Theta_{D^*}$ is defined as the angle between the $\pi_s$ and the
direction of \Dstr in the \Dstr rest frame. Similarly $\Theta_{B}$ is
the angle between the $\pi_f$ and direction of the \B in the \B rest
frame. Since the \B meson is a pseudo-scalar its decay is isotropic. We
thus expect the $\cos(\Theta_{B})$ distribution to be flat for
signal. Since the \Dstr is longitudinally polarized along its direction
of travel when measured from its rest frame we expect the
$\cos(\Theta_{D^*})$ distribution to be distributed as
$\cos^2\Theta_{D^*}$ in signal.

Figure \ref{fig_dstrpi-part} shows projections of a 2-dimensional
 $\chi^2$ fit to the data. The fitting function consists of shapes for
 signal, continuum and \bbar background. The signal and \bbar background
 components were determined from Monte Carlo simulations while the
 continuum shape was modeled by off-resonance data. The overall
 normalization of the \bbar background component is allowed to vary in
 the fit while the continuum component is fixed. The fitting function
 used in the $\pi^-_f\pi^0$ fit contains an additional component to
 accommodate the contribution of the \Dstrpi decay where the \Dstrp
 decays to $\Dp\piz.$ This is fixed by the \Dstrpi branching fraction
 determined in the $\pi^-_f\pi_s^+$ fit. The measured branching
 fractions are:
$$
\begin{array}{lcl}
{\cal B}(\Dstrpi)&=&(2.81\pm 0.11\pm 0.21\pm 0.05) \times 10^{-3} \\
{\cal B}(\Dstrzpi)&=&(4.34\pm 0.33\pm 0.34\pm 0.18) \times 10^{-3} 
\end{array}
$$
where the first error is statistical the second is systematic and the
third is due to the uncertainty in the \Dstr branching fractions.
These results are in excellent agreement with the results obtained by
the full reconstruction technique. 

\begin{figure}[t]
\centerline{\epsfxsize 12.0 truecm\epsfbox{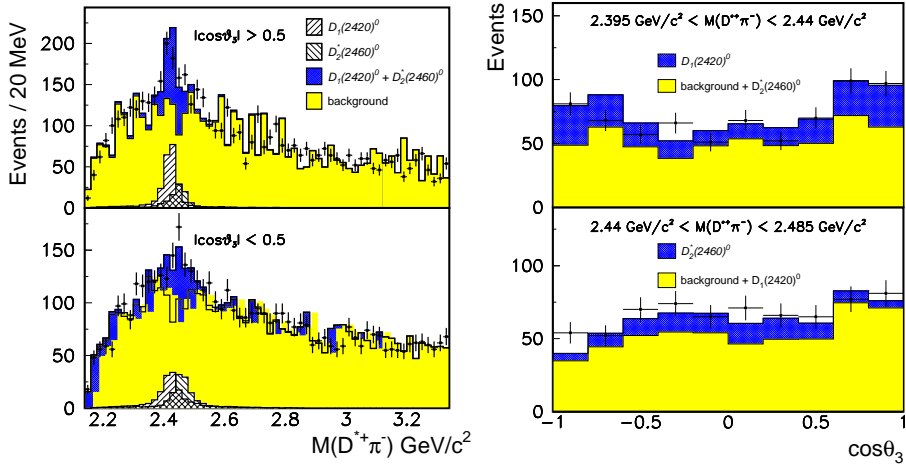}}
\caption{\label{fig_ddblstrpi} Projections of 2-D fit to data for the
variables $\cos\theta_3$ and $m_{D^{*+}\pi^-}$. The two figures on the
left show the $m_{D^{*+}\pi^-}$ in different regions of the $\cos\theta_3$. 
The figures on the right show the projection of $\cos(\theta_3)$ in
different regions of $m_{D^{*+}\pi^-}$. The data are the points with
error bars.} 
\end{figure}

\subsection{$B \to D_J^0(n\pi)^-$ decays}
The full and partial reconstruction techniques have been used by both
CLEO and ARGUS to reconstruct decays of the \B meson to excited $L=1$
charmed mesons. Published results exist only for the $B^- \to
D_1^0(2420)\pi^-$ mode. For the $B^-\to D_J^0\rho^-$ mode upper limits from
\cleo~\cite{BigB} are available for the $B^-\to \Ddbhighz\rho^-$ and
$B^\to \Ddblowz\rho^-$ decays and from ARGUS a measurement of the sum
over all $D^0_J$ has been reported~\cite{PDG96}.

There are four excited states $D_J^{(*)}$ mesons with in a $L=1$
orbital angular momentum state. Two, the $D_2^*(2460)$ and the
$D_1(2420)$ are narrow resonances which have been seen and their decays
measured~\cite{PDG96}. Angular momentum and parity conservation
place restrictions on the strong decay of these states. The
 $D_2^*(2460)$ can decay via D-wave to either $D\pi$ or $\D^*\pi$ while
the $D_1(2420)$ can decay to $\D^*\pi$ via S-wave or D-wave. 

Recently, \cleo has reported new measurements on the $\Ddblowz\pim$ and
$\Ddbhighz\pim$~\cite{Ddblconf} using a partial reconstruction
technique.  The analysis follows the procedure outlined in
Section~\ref{Partial}. Once again the reconstruction of the decay
depends on knowing the masses of the decay products, the beam-energy and
the 4-momenta of the three pions, $\pi_1,\pi_2,\pi_3$ produced in the
decay chain $B\to D_J^0\pi_1, D_J^0\to D^{*+}\pi^-_2$ and $\Dstr \to
D^+\pi_3$. In this analysis the mass of the $\Dstrp\pim$ combination and
the helicity angular distribution $\cos\theta_3$ are used to identify
the signal. The variable $\theta_3$ is defined as the angle between the
\Dz and the direction of the \Dstr in the \Dstr rest frame and describes
the helicity angular distribution of the \Dstr in the decay chain
$\DJ\to D^*\pi_2,D^*\to D\pi_3$. From angular momentum conservation in
$B\to \Ddbhigh\pi$ we expect that the $\cos\theta_3$ distribution to be
proportional to $\sin^2\theta_3$ while in $B\to \Ddblow\pi$ we expect the
distribution to be proportional to $1+3\cos^2\theta_3$~\cite{Ddblconf}.

Figure \ref{fig_ddblstrpi} shows the projections of 2-dimensional
unbinned likelihood fits to the data. The two figures on the left show
the projection onto the $m_{D^{*+}\pi^-}$ axis in different regions of
$\cos\theta_3$ where either the \Ddblow or the \Ddbhigh are expected to
dominate. The long flat tails in the signal shape result from
combinations where a random track has faked a real $\pi_2$. The two plots
on the right show the projection of fit on the $\cos\theta_3$
distribution. Here too, the regions where either the \Ddblow or \Ddbhigh
are expected to dominate are separated by $m_{D^{*+}\pi^-}$ cuts. The
signal and \bbar background shapes were determined from Monte Carlo
simulations. The continuum component was modeled in real data taken
below the \USSSS peak. The measured product branching fraction are:
$$
\begin{array}{lcl}
{\cal B}(B^-\to \D_1^0\pi^-)\times{\cal B}(\D_1^0\to D^{*+}\pi^-)      &=&(7.8\pm 1.6\pm 1.0\pm 0.2)\times 10^{-4} \\
{\cal B}(B^-\to \D_2^{*0}\pi^-)\times{\cal B}(\D_2^{*0}\to D^{*+}\pi^-)&=&(4.2\pm 1.6\pm 0.6\pm 0.1)\times 10^{-4} \\
\end{array}
$$
The branching fraction for the individual \B decays are listed in Table
\ref{tbl_dnpi} where the authors have used isospin conservation and the measured
branching fraction of $\Ddbhighz\to \Dp\pim$~\cite{ddblstrmeas} to deduce
the following $D_J$ decay rates 
$$
{\cal B}r(\Ddblowz\to\Dstrp\pim)=2/3 \ {\rm and} \ {\cal B}r(\Ddbhigh\to\Dstrp\pim)=0.20
$$
The authors have neglected multi-pion decays to obtain these branching
fractions.  

\begin{table}
\begin{center}
\caption{\label{tbl_DDs}Branching Fractions for $B\to D^{(*)}D_s^{*+}$ Decay Modes~\protect\cite{DDs,changes}}
\begin{tabular}{|ll||ll|}\hline 
  Mode 	       		&\mco{1}{c||}{${\cal B}r$ (\%)}& Mode		&\mco{1}{c|}{${\cal B}r$ (\%)}\\ \hline
 $D^+D_s^-$		& $0.87\pm 0.24\pm 0.22$ & $D^0D_s^-$		& $1.18\pm 0.21\pm 0.27$ \\
 $D^+D_s^{*-}$		& $1.00\pm 0.35\pm 0.25$ & $D^0D_s^{*-}$	& $0.82\pm 0.25\pm 0.19$ \\
 $D^{*+}D_s^-$		& $0.87\pm 0.22\pm 0.18$ & $D^{*0}D_s^-$	& $1.32\pm 0.40\pm 0.37$ \\
 $D^{*+}D_s^{*-}$	& $1.91\pm 0.47\pm 0.41$ & $D^{*0}D_s^{*-}$	& $2.91\pm 0.83\pm 0.70$ \\ \hline
\end{tabular}					      
\end{center}
\end{table}					      

\subsection{$B \to D^{(*)}D^{(*)}_s$ decays}
Another important class of two-body decays are the decays to two charmed
mesons. The Cabbibo allowed process produces either a $D_s^{*+}$ or a
 $D_s^+$ and can proceed only via the external spectator diagram. The
large mass of the $D^{(*)}D_s^{(*)+}$ system implies a lower momentum
transfer and hence provides a means by which to probe a different $q^2$
region than is possible with $D^{(*)+}(n\pi)$ decays. Measurements of
 $B\to D_s^{(*)+}D^{(*)}$ rates together with measurements on the $B\to
D^{*+}\pi^-,\rho^-$ rates allow the extraction of the $f_{D_s}$ and
$f_{D^*_s}$ decay constants~\cite{NS97}.

Both CLEO and Argus~\cite{DDs,PDG96} have measured \B decays to
Cabbibo allowed double charm final states. The \cleo analysis reconstructs 
all eight decay modes exclusively using several the three $\Dz,$ one \Dp
and several $D_s$ sub-channel decays. Only a subset consisting of 
$2.04 fb^{-1}$ of the complete \cleo on resonance data were used for this
analysis. The values listed in \tbl~\ref{tbl_DDs} are the published
\cleo values rescaled by the \D and \Dstr branching fractions in
Ref. [8]. 

\subsection{$B^- \to D^0K^-$ decays}
A search for the Cabbibo suppressed decay to $D^0K^-$ has been performed
by \cleo\cite{DzKconf}. Interference between the $b\to c\bar{u}s$ and 
 $b\to u\bar{c}s$ decay, which hadronize as $B^- \to D^0K^-$ and $B^- \to
\bar{D^0}K^-$ respectively can be used to determine the CKM phase
($\gamma$)~\footnote{This topic is discussed in greater detail in an
article by D. Atwood in these proceedings.}. 

\begin{figure}
\centerline{\epsfxsize 12.0 truecm\epsfbox{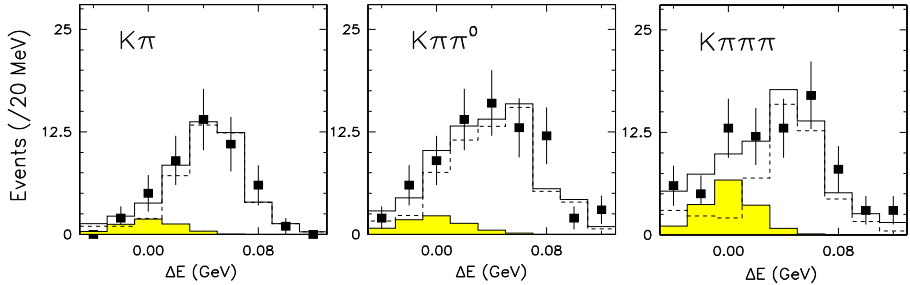}}
\caption{\label{fig_D0K} The distribution of the reconstructed energy
minus the beam-energy for candidates the satisfy the full reconstruction
selection algorithm. The data are the points, the solid hatched
histogram is the signal the dashed histogram is assumed background,
taken from the mass sidebands and the fit is shown by the open
histogram. From left to right the $D^0\to K^-pi^+,K^-\pi^+\pi^0$ and
$K^-3\pi$ modes.}
\end{figure}

The analysis procedure features a full reconstruction technique but
\ediff is used extract the event yield. For this analysis particle ID is
significantly more important because of the large background from the
Cabbibo allowed \Dzpi process. At CLEO, the current particle ID system
cannot distinguish with great certainty pions from kaons at high
momentum. For example $K\pi$ separation at 2.2 \pGeV is less than
$2\sigma$ where $\sigma$ is the difference between the expected and
measured energy loss $dE/dx$ due to specific ionization in the drift
chamber.  \fig~\ref{fig_D0K} shows the \ediff distribution for the
$B^-\to D^0K^-$ with $D^0\to K^-\pi^+,K^-\pi^+\pi^0,\K^-\pi^+\pi^-\pi^+$
combinations consistent with \B production. The large background from
misidentified \Dzpi is seen to overwhelm the signal. The limit reported is 
$$
{\cal B}r(B^-\to D^0K^-) < 4.4 \times 10^{-4} \ @\ 90\%\ {\rm C.L.}
$$

\section{Test of Factorization}
Theoretical models~\cite{NS97,BSW,NRSX,Deandrea,ReaderIsgur} of hadronic
decays of heavy mesons invoke the factorization approximation to make
definite predictions on decay rates to exclusive modes. Factorization is
used in order to reduce the hadronic matrix elements to products of
factorized matrix elements with one describing the creation of a hadron
from the vacuum and the other describing the transition of the \B
meson. The factorized $\B\to D$ transition matrix element is equivalent
to the matrix element encountered in semi-leptonic decays while the
matrix element describing the creation of a meson from the vacuum is
parameterized by the decay constant of the meson~\cite{NS97}.

Some theoretical motivation exists, at least for decays with large
energy release, for the factorization hypothesis. These are based on
``color transparency arguments'' which postulate that a $q\bar{q}$ pair
created in a point-like interaction will hadronize only after a time
given by its $\gamma$ factor times a typical hadronization
scale~\cite{NS97}. Thus in an energetic transition the hadronization of
a light $q\bar{q}$ pair which travels together in the same direction
will not occur until it is a significant distance from the
interaction region~\cite{NS97}. While this scenario describes decays
such as $B\to D^{(*)}(n\pi)^-$  where $n\pi$ is a light meson it does
not apply to $B\to D^{(*)}D_s^{(*)}$ decays where the $\gamma$ factors
are smaller. It remains to be seen if the factorization hypothesis holds
for decays that occur at lower energy transfers.

\def\diffbr{\frac{d{\cal B}r}{dq^2}(\bar{B}\to D^{*}l\nue) }
\subsection{Branching Fraction Tests}
To test the factorization approximation we exploit the similarity
between hadronic two-body decays and semi-leptonic decays. In
semi-leptonic decays factorization is strictly obeyed since the 
leptonic current does not interact with the hadronic part. By taking the 
ratio of a hadronic decay rate to its semi-leptonic counterpart we can
compare the experimental ratio to the theoretical expectation
and thus performing a direct test of the factorization
hypothesis~\cite{NRSX}. If Factorization holds then 
$$
\frac{{\cal B}r(\bar{B^0}\to D^{*+}h^-)}{\diffbr |_{q^2=m_h^2}}=
6\pi^2f_h^2|a_1|^2|V_{ij}|^2X_h 
$$
is satisfied. The branching fraction measurements listed in 
\tbl~\ref{tbl_dnpi} together and measurements of the semi-leptonic rate at
the appropriate $q^2$ are used. The value of the QCD parameter $a_1$ is
taken to have the strict factorization value $a_1 = c_1(\mu) +
\frac{1}{N_c}c_2(\mu)= 1.04$, where $c_1(\mu)$ and $c_2(\mu)$ are the
Wilson coefficients calculated at $\mu=m_b$ and $1/N_c=1/3$~\cite{NS97}. 
The semi-leptonic branching fraction is interpolated~\cite{TEBreview} from
a fit to the differential branching fraction at the appropriate
 $q^2$. The $X_h$ is a kinematic factor which depends
on the masses of the hadrons and relevant form factors and is close to
1.0. Table~\ref{tbl_fact} shows a comparison between the
experimental and the theoretical ratios for different regions of
 $q^2$. The agreement between data and theory is quite good in the low
$q^2$ region suggesting that factorization works, at least for energetic
Class I decays.

\begin{table}
\begin{center}
\caption{\label{tbl_fact}Test of Factorization: Comparison of $R_{\rm
exp}$ to $R_{\rm theo}$~\protect\cite{Fact_inputs,changes}}
\begin{tabular}{|lll|}\hline
\mco{1}{|c}{$q^2$}	&\mco{1}{c}{$R_{\rm exp} (\rm GeV^2)$}
&\mco{1}{c|}{$R_{\rm theo} (\rm GeV^2)$}\\ \hline  
 $m_{\pi}^2$ 	& $1.13\pm 0.04\pm 0.14$& $1.09\pm 0.07$ \\
 $m_\rho^2$  	& $2.94\pm 0.24\pm 0.48$& $2.68\pm 0.20$ \\
 $m_{a_1}^2$ 	& $3.45\pm 0.36\pm 0.59$& $3.19\pm 0.33$ \\
 $m_{D_s}^2$ 	& $1.81\pm 0.45\pm 0.40$& $3.50(f_{D_s}/240{\rm MeV})$ \\
 $m_{D_s^*}^2$ 	& $3.76\pm 0.93\pm 0.84$& $3.21(f_{D^*_s}/275{\rm MeV})$ \\
 \hline
\end{tabular}
\end{center}
\end{table}

\section{The Relative Sign and Amplitude of $a_1$ and $a_2$}
Theoretical models based on the BSW approach relegate all
short-distance QCD effects into QCD parameters~\cite{BSW,NRSX,Deandrea}. 
In two-body tree-level decays, two parameters, $a_1$ and $a_2$ multiply
the dominant spectator contribution from the external and internal
diagrams, respectively. They provide important clues into the role
played by the strong interaction. For instance, the absolute value of the QCD
parameters are sensitive to the factorization scale and additional
long-distance contribution~\cite{NS97}. The latter being particularly
true for the $a_2$ since it involves the difference between two small
numbers. Also, the relative sign of $a_1$ and $a_2$ identifies the kind
of interference between the internal and external spectator diagrams. 

To determine the values of $a_1$ and $a_2/a_1$ we use the
branching fraction measurements in \tbl~\ref{tbl_dnpi} and \ref{tbl_DDs}
together with the theoretical predictions from the Neubert \etal\ model 
calculations~\cite{NRSX,NS97}. Since the QCD parameters are 
expected to be process independent~\cite{NS97}, at least for decays
that occur at similar momentum scales, we perform a least squares to
four of the  $D^{(*)}(n\pi)$ decays, and a separate fit to the
$D^{(*)}D_s^{(*)+}$ decays. The results are:
$$
\begin{array}{lcl}
|a_1|_{Dn\pi} &=& 1.03 \pm 0.02 \pm 0.04 \pm 0.05 \\
|a_1|_{DD_s}  &=& 1.01 \pm 0.05 \pm 0.12 \pm 0.05
\end{array}
$$
where the first error is statistical, the second is the systematic
error~\footnote{The first systematic error of the $|a_1|_{DD_s}$ is
computed assuming all systematic errors except the error from the 
${\cal B}r(D_s^+\to\phi\pi^+)$, are independent in the fit. This 
error is a common systematic and is thus added in quadrature.}
and the third is the error due to the uncertainty in the \B
lifetime/production ratio~\cite{Dstrlnu}. Consistent
results are obtained using the Deandrea \etal\ model and the Nuebert and
Stech ``New Model'' in Ref. [1]. No process dependency is
observed in the data.

To determine the value and the relative sign $a_2/a_1$ we form ratios of
Class III to Class I branching fractions; again using the Neubert \etal\ model
calculations to extract the value from a least squares fit to data. The
value obtained from the fit was:
$$
{a_2\over a_1} = +0.21 \pm 0.03 \pm 0.03 \mp^{0.13}_{0.12}
$$
The lifetime-production fraction is taken to be 1.0 in both the fit to
$a_1$ and $a_2/a_1$. This is consistent with the \cleo
measurement~\cite{Dstrlnu} $\frac{f_{+}\tau_{+}}{f_{0}\tau_{0}} =1.14
\pm 0.14 \pm 0.13.$  

The positive sign for $a_2/a_1$ suggests that the internal and external
decay amplitudes interfere constructively. This is in contrast to the
situation in $D$ decay where a negative sign implies destructive
interference.  The destructive interference in $D$ decays is responsible
for the longer lifetime of the charged $D$ mesons which can proceed only
through the external diagram. The positive sign of $a_2/a_1$ seems to
suggest a shorter lifetime for the charged \B which is not observed in
data. One possible explanation for this may be that constructive
interference is only found in low multiplicity \B decays which
constitute a small fraction of the total hadronic width. It remains to
be seen if this pattern persists for other \B decays.

\section{Conclusions}
Significant improvements in the precision of branching fraction
measurements of exclusive hadronic \B decays have been made in the
recent past. This is primarily due to the ever increasing statistics,
particularly at \bbar threshold machines, and improvements in the analysis
procedure. From these precise measurement we are now able to more
accurately test theoretical predictions based on factorization. We
showed in \tbl~\ref{tbl_fact} a comparison between the experimental
results and theoretical predictions based on factorization. We find that
in energetic Class I transition the factorization hypothesis seems to be
well supported by the data. It remains to be seen if factorization holds
in the higher $q^2$ region and of the color-suppressed $D^{(*)0}n\pi$
decays. So far none of these color-suppressed decays have been
observed. Also, the QCD parameters $a_1$ and the $a_2/a_1$ have been
determined by fitting the data to model calculations.  In Class III the
relative sign of $a_2/a_1$ was found to be positive indicating that the
interference between internal and external decays diagrams is
constructive. This fact provides additional insight into differences
between the $D$ and $B$ decays which occur at different factorization
scales.

\section{Acknowledgments}
I would like to thank those who participated or contributed to these
analyses, in particular the \cleo, and OPAL collaborations. I would also
like to take this opportunity to thank Tom E. Browder and Sandip Pakvasa
and their staff for organizing an excellent Conference with a stimulating
scientific program.


\begin{thebibliography}{99}
\bibitem{pfave} This numbers represents a weighted average of recent
\cleo results obtained from the full and partial reconstruction
analyses described in Refs. [8] and [14]. Since
the measurement were conducted on the same experiment some of the
systematic errors are correlated. In averaging the results these
correlations were taken into account using the error matrix.
\bibitem{nrrho} The non-resonant $\pi^0\pi^-$ component in
$D^{(*)}\rho^-$ is small. It ranges from less than 2.5\% to 9.0\%
depending on the mode. For these modes we neglect non-resonant
contributions. This analysis was performed in reference~\cite{BigB} on
approximately 1/3 of the current data sample.
\bibitem{nra1} The non-resonant $\pi\pi\pi$ and $\pi\rho^0$ component in
$D^{(*)}a_1^-$ was found to be less than 12\% and 24\% in $D^0a_1$ and
$D^+a_1$ decays at the 90\% Confidence Level~\cite{JRthesis}. The event yields 
were corrected accordingly for the $D^{(+,0)}a_1^-$ modes. In $D^*a_1$
the contributions from non-resonant $\pim\rhoz$ and $D^{**}\rho$ were
found to be less 9.4\% and 10.6\% respectively at the 90\% confidence
level~\cite{BigB}. We thus neglect non-resonant contributions to both
$D^{*(+,0)}a_1^-$ modes.
\end{thebibliography}
\end{document}